\documentclass[conference]{IEEEtran}
\IEEEoverridecommandlockouts

\usepackage{amsmath,amssymb,amsfonts}
\usepackage[T1]{fontenc}
\usepackage{graphicx}
\usepackage{textcomp}
\usepackage{booktabs}
\usepackage{multirow}
\usepackage{enumitem}
\usepackage{subcaption}
\usepackage{longtable}  
\usepackage{xcolor}
\usepackage{tikz}
\usepackage{cuted}

\usepackage{svg}

\usepackage[
 backend=biber
,style=ieee
,minbibnames=1
,maxbibnames=2
,mincitenames=1
,maxcitenames=2
,eprint=true
,doi=true
,isbn=false
,url=true
]{biblatex}
\bibliography{ref} 
\AtBeginBibliography{\footnotesize}

\DeclareSourcemap{
  \maps{
    \map{
      \step[fieldset=month, null]
      \step[fieldset=address, null]
      \step[fieldset=location, null]
      \step[fieldset=publisher, null]
      \step[fieldset=url, null]
      \step[fieldset=isbn, null]
      \step[fieldset=editor, null]
    }
  }
}

\usepackage{tikz}

\usepackage{algorithm}
\usepackage[noend]{algpseudocode}
\algnewcommand{\LineComment}[1]{\State \(\triangleright\) #1}
\algnewcommand{\algorithmicand}{\textbf{ and }}
\algnewcommand{\algorithmicor}{\textbf{ or }}
\algnewcommand{\OR}{\algorithmicor}
\algnewcommand{\AND}{\algorithmicand}
\algnewcommand{\var}{\texttt}

\usepackage[
hidelinks,
  colorlinks=true,
  citecolor=blue,
  pdfstartview=Fit,
  breaklinks=true
]{hyperref}

\usepackage{cleveref}
\Crefname{figure}{Fig.}{Figs.}

\usepackage{flushend}

\begin{document}

% \title{Inter-Flow Service Degradation Detection:\\ An Analysis of Opportunities and Challenges}
\title{On the Feasibility of Inter-Flow Service Degradation Detection}

\author{
    \IEEEauthorblockN{Balint Bicski\IEEEauthorrefmark{1}\IEEEauthorrefmark{3} and Adrian Pekar\IEEEauthorrefmark{1}\IEEEauthorrefmark{2}\IEEEauthorrefmark{3}}
    \IEEEauthorblockA{
        \IEEEauthorrefmark{1}
        % Department of Networked Systems and Services, Faculty of Electrical Engineering and Informatics,\\ 
        Budapest University of Technology and Economics, M\H{u}egyetem rkp. 3., H-1111 Budapest, Hungary.\\
        \IEEEauthorrefmark{2}HUN-REN-BME Information Systems Research Group, Magyar Tud\'{o}sok krt. 2, 1117 Budapest, Hungary.\\
        \IEEEauthorrefmark{3}CUJO LLC, Budapest, Hungary.\\
        Email: balint.bicski@cujo.com, apekar@hit.bme.hu
    }
}

\maketitle

\begin{abstract}
Hardware acceleration in modern networks creates monitoring blind spots by offloading flows to a non-observable state, hindering real-time service degradation (SD) detection. To address this, we propose and formalize a novel inter-flow correlation framework, built on the hypothesis that observable flows can act as environmental sensors for concurrent, non-observable flows.
We conduct a comprehensive statistical analysis of this inter-flow landscape, revealing a fundamental trade-off: while the potential for correlation is vast, the most explicit signals (\textit{i.e.}, co-occurring SD events) are sparse and rarely perfectly align. Critically, however, our analysis shows these signals frequently precede degradation in the target flow, validating the potential for timely detection. We then evaluate the framework using a standard machine learning model. While the model achieves high classification accuracy, a feature-importance analysis reveals it relies primarily on simpler intra-flow features. This key finding demonstrates that harnessing the complex contextual information requires more than simple models. Our work thus provides not only a foundational analysis of the inter-flow problem but also a clear outline for future research into the structure-aware models needed to solve it.
\end{abstract}


\begin{tikzpicture}[remember picture,overlay]
\node[anchor=north, align=center, text=black, font=\small\itshape, yshift=-.6cm] at (current page.north) {
    This version of the paper has been accepted for presentation at the 5th International Workshop on 
    Analytics for Service and Application\\ Management (AnServApp 2025),
    co-located with the 21st International Conference on Network and Service Management (CNSM 2025)
};
\end{tikzpicture}

\vspace*{-.6cm}

\section{Introduction}

% Service degradation (SD) detection in modern network infrastructure faces the fundamental challenge of balancing comprehensive monitoring with resource constraints. Contemporary network devices employ dual-path architectures where flows transition from CPU-based processing (observable state) to hardware-accelerated forwarding (non-observable state) to maintain performance~\cite{michel2017era}. This architectural necessity creates monitoring blind spots where traditional packet-level analysis becomes impossible.

Service degradation (SD) detection in modern network infrastructure faces a fundamental trade-off between comprehensive monitoring and resource constraints. To achieve line-rate speeds, contemporary network devices employ dual-path architectures where network flows transition from CPU-based processing (the \textit{observable state}) to specialized hardware for accelerated forwarding (the \textit{non-observable state})~\cite{michel2017era}. This transition, while essential for performance, offloads the flow from software visibility, creating a critical \textit{monitoring blind spot} where traditional packet-level analysis becomes impossible.

% Our previous research established methodologies for detecting SD using LAN-side delay analysis~\cite{bicski2024unveiling} and demonstrated the feasibility of \textit{intra-flow} SD prediction, where a flow's own early packets are used to predict its future degradation~\cite{bicski2024early}. While valuable, this intra-flow approach is predictive by nature; it cannot provide \textit{immediate detection} for flows already in their non-observable state, a critical limitation for real-time network management.

Our previous research established the feasibility of \textit{intra-flow} SD prediction, where a flow's own early characteristics are used to predict the likelihood of its future degradation~\cite{bicski2024unveiling, bicski2024early}. While valuable, this approach is fundamentally predictive---yielding only a probability of future issues---and cannot provide \textit{immediate detection} for flows already in their non-observable state, a critical limitation for real-time network management.

This paper proposes an \textit{inter-flow} (or cross-flow) detection method that addresses this gap. Our approach is founded on the principle that network SD often manifests as a system-wide phenomenon, impacting multiple concurrent flows~\cite{Valdez2020,smolyak2020mitigation}. We hypothesize that by leveraging the temporal correlations between concurrent flows, we can use those in an observable state as environmental sensors to infer the real-time status of flows that are already non-observable.

% This paper proposes an \textit{inter-flow} (or cross-flow) detection method that addresses this gap. Our approach is founded on the principle that network service degradation often manifests as a system-wide phenomenon, impacting multiple concurrent flows~\cite{Valdez2020,smolyak2020mitigation}. The concept of leveraging temporal correlations between different system elements to diagnose system-wide issues is well-established in fields like multi-sensor systems and temporal network analysis~\cite{li2022correlation, Saqr2024}. We hypothesize that such correlations exist between concurrent network flows. 

To test this hypothesis, our work unfolds in two stages. First, we formalize the inter-flow framework and conduct a deep statistical characterization of the underlying data landscape to validate that the potential for correlation is practically significant and timely. Second, we perform a preliminary experimental evaluation by implementing a detection model using a straightforward feature construction method. This allows us to establish a crucial first performance benchmark and, more importantly, to uncover the practical challenges of harnessing this complex, high-dimensional data.

Our main contributions are: 
\begin{enumerate}
    \item The formalization of a novel \textit{inter-flow correlation framework} to address the hardware acceleration monitoring gap. 
    \item A comprehensive \textit{statistical analysis} that characterizes the inter-flow landscape and validates the potential for timely detection.
    \item A preliminary \textit{experimental evaluation} that establishes a performance benchmark and uncovers the limitations of using a simple feature \textit{concatenation} strategy for this task. 
    \item A clear \textit{roadmap for future research}, emphasizing the need for structure-aware models to unlock the full potential of inter-flow analysis.
\end{enumerate}
Additionally, we complement these contributions with a comprehensive \textit{digital artifact} containing detailed exploratory data analysis, feature engineering pipeline, and extended results.

The remainder of this paper is organized as follows: \Cref{sec:related} reviews related work. \Cref{sec:methodology} introduces our framework. \Cref{sec:inter_flow_stats} presents the statistical analysis of the inter-flow landscape. \Cref{sec:evaluation} evaluates the detection model and discusses its limitations. Finally, \Cref{sec:conclusion} concludes with future work directions.

\section{Related Work}
\label{sec:related}

While extensive research exists on intra-flow analysis and general network correlation, the specific challenge of using observable flows to detect service degradation in concurrent non-observable flows remains largely unaddressed. Our work builds upon several distinct research areas while addressing this novel intersection.

\subsection{Intra-Flow Service Degradation Detection}
The foundation of our work is our previously established methodologies for flow-based SD detection. We define SD as statistically significant increases in LAN-side packet delay and jitter, identified using methods from~\cite{bicski2024unveiling}. Building on this, our prior work demonstrated the feasibility of \textit{intra-flow SD prediction}~\cite{bicski2024early}. In that approach, statistical features (e.g., mean and variance of delay) from a flow's first few observable packets were used to predict the likelihood of future degradation in the same flow, achieving a balanced accuracy of 0.84. However, its critical limitation---and the primary motivation for this paper---is that it is a \textit{predictive} tool, not a real-time detection method. This prior work establishes the clear research gap we aim to fill and serves as the performance benchmark for our new inter-flow strategy.

\subsection{Inter-Flow and Cross-Correlation Analysis}
The limitations of single-flow analysis motivate exploring inter-flow dynamics. The concept of using correlation in networking is not new, but has been applied to different problems. For instance, some have correlated control and data plane traffic for security anomaly detection~\cite{assadhan2020network}. While effective, their focus on inter-plane correlation is distinct from our flow-to-flow approach. More recently, frameworks like FlowID have used hypergraphs to model inter-flow relationships for the purpose of traffic classification~\cite{zhou2025multiview}. These studies validate the utility of analyzing relationships between flows, but their objective is classification, not the real-time detection of performance degradation.

\subsection{Conceptual Parallels in Temporal Systems}
The principle of leveraging temporal correlations is well-established in other complex domains, providing a strong conceptual parallel to our work. In multi-sensor systems, for example, researchers have constructed temporal correlation graphs between different sensor readings to treat anomaly detection as a graph classification problem, achieving F1-scores exceeding 0.90~\cite{li2022correlation}. Just as they leverage correlations between heterogeneous sensors to diagnose system-wide anomalies, our work leverages correlations between concurrent network flows to infer the health of the network. This supports our hypothesis that such an approach can reveal degradation patterns that are invisible to single-element analysis~\cite{Saqr2024}.

\subsection{The Research Gap: Monitoring in Accelerated Networks}
These distinct areas of research converge on a critical, unaddressed problem: the monitoring blind spot created by hardware acceleration in modern networks~\cite{firestone2018azure}. Once a flow is offloaded to a hardware path, its real-time performance metrics become invisible to standard monitoring tools like NetFlow/IPFIX~\cite{rfc7011,rfc3176}. While the techniques above have established intra-flow prediction and explored inter-flow correlation for other tasks, to our knowledge, no prior work has specifically aimed to bridge this hardware-induced monitoring gap by using concurrent observable flows to infer the real-time degradation state of non-observable ones. This is the novel contribution of our paper.

\section{Methodology}
\label{sec:methodology}

This section details the framework for our inter-flow analysis. We first explain the core problem of monitoring in hardware-accelerated networks, then define our O/NO segmentation framework, and finally, formalize the temporal correlation approach used to overcome the monitoring challenge.

\subsection{The Hardware Acceleration Blind Spot}
Modern network devices, such as residential gateways, rely on hardware acceleration to achieve high-speed packet forwarding. In this common architecture, a network flow’s lifecycle is split across two processing paths. When a new flow begins, its first few packets are processed on the device's main CPU. In this \textit{Observable (O) state}, software has full visibility, allowing for deep packet inspection and the measurement of performance metrics like per-packet delay.

To maintain line-rate performance and reduce CPU load, the device's software programs the hardware forwarding engine (the ASIC or ``fast path'') with a rule to handle all subsequent packets of that flow. The flow then transitions to the \textit{Non-Observable (NO) state}. In this state, packets bypass the CPU entirely and are processed directly in hardware. This creates a critical \textit{monitoring blind spot}: while the flow is active and carrying user traffic, the software has no visibility into its real-time performance. Our work is designed specifically to infer the performance of a flow during this non-observable phase.

\subsection{The O/NO Segmentation Framework}
Our work formalizes this process as the O/NO flow segmentation framework, building upon our previous work~\cite{bicski2024early}. The transition from the O to the NO state occurs after a predefined number of a flow's initial delay measurements, $m$, have been observed. This study uses O/NO splits of $m \in \{5, 10\}$. The inter-flow approach presented here aims to shift from the predictive capabilities of our prior work to a real-time detection method for flows in the NO state.

\subsection{Inter-Flow Correlation Principle}
The core assumption driving our methodology is that SD often manifests as a system-wide phenomenon, impacting multiple concurrent flows. When network infrastructure experiences congestion or other bottlenecks, these conditions create detectable temporal correlations between the performance characteristics of flows. Our framework leverages this principle, using concurrent observable flows as distributed environmental ``sensors'' to infer the state of a non-observable target flow.

\begin{figure}[ht]
    \centering
    \includegraphics[width=\columnwidth]{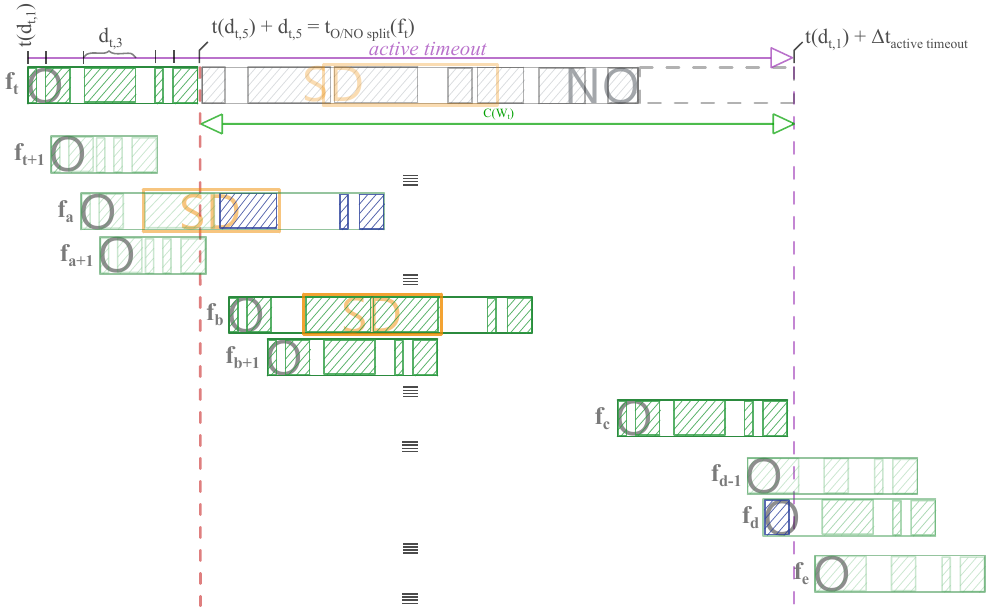}
    \caption{The inter-flow detection concept: leveraging the observable delay measurements from concurrent flows to infer the state of a target flow in its non-observable phase.}
    \label{fig:inter_flow_concept}
\end{figure}

\Cref{fig:inter_flow_concept} provides a detailed illustration of this concept. A target flow, $f_t$, enters its non-observable state. The goal is to infer its status by analyzing the features of other flows ($f_a, f_b, \dots, f_d$) whose observable segments are active during the non-observable lifetime of $f_t$.

\subsection{Temporal Correlation Framework}
We formalize this relationship by defining a \textit{Correlation Window} $W_{t}$ for each target flow $f_t$. This window begins the moment $f_t$ transitions to its non-observable state and has a duration $\Delta t = \textit{active timeout} - \Delta t_{\textit{O}}$, where $\Delta t_{\textit{O}}$ is the time the target flow spent in its observable state. This definition ensures the window covers the maximum possible lifetime of the flow's non-observable segment before it expires from the flow cache.

A \textit{covering flow}, $f_k$, is any flow that has at least one observable delay measurement within $W_t$. To specify this formally, we first define $F_O(W_t)$ as the set of all flows that are in their observable state at some point during the window $W_t$, and $F_{NO}(W_t)$ as the set of flows in their non-observable state. The correlation space $C(W_{t})$ is then defined as the set of all pairs $(f_t, f_k)$ that meet the temporal overlap conditions specified in \Cref{eq:cw1,eq:cw2}. Here, $d_{x,y}$ represents the value (duration) of the $y^{th}$ delay measurement of flow $f_x$, while $t(d_{x,y})$ denotes the timestamp marking the beginning of that delay measurement.

{\scriptsize
    \begin{align}
    \label{eq:cw1}
        C(W_{t}) = \{(f_{t}, f_k) : f_{t} \in F_{NO}(W_{t})\ \land\ f_k \in F_O(W_{t})\} \\[1ex]
    \label{eq:cw2}
        \forall k\ \exists i \in \{1,\ 2,\ ...,\ m\}\ \land\ \exists j \in \{1,\ 2,\ ...,\ m\}: \\
        i \le j \; \land \;  t(d_{t,m}) + d_{t,m} \le t(d_{k,i}) \; \land \;  t(d_{k,j}) + d_{k,j} \le t(d_{t,0}) + \Delta t \notag
    \end{align}
}

This general framework is designed to be comprehensive. As \Cref{fig:inter_flow_concept} illustrates, it accommodates all types of temporal overlaps. For instance, some covering flows like $f_a$ and $f_d$ only \textit{partially overlap} the correlation window. In these cases, our framework considers only the delay measurements that fall within the window (highlighted in blue). Other flows, like $f_b$ and $f_c$, are \textit{fully contained} within the window. The framework also accounts for \textit{non-contributing} flows, such as $f_{t+1}$ and $f_e$, which are active but have no observable delay measurements inside $W_t$.
% This section defines the complete universe of potential data; the subsequent sections will explore the properties of this data and test a specific, practical implementation.

\section{Inter-Flow Coverage Statistics}
\label{sec:inter_flow_stats}

Before building a detection model based on the framework in \Cref{sec:methodology}, it is essential to first validate its core assumption: \textit{does the necessary data for inter-flow correlation actually exist in a real-world setting, and is it timely enough to be useful?} This Section addresses these questions by presenting a statistical characterization of the inter-flow landscape. We applied our framework to a large dataset of network flows, using an O/NO split of 10, to characterize the practical properties of the covering flows.

The following analysis is primarily descriptive, designed to characterize the raw data landscape and validate our core assumptions before moving to a predictive modeling phase. Consequently, rather than focusing on specific correlation coefficients such as Pearson or Spearman---which test for linear or monotonic relationships between variables---our analysis centers on the empirical distributions of direct counts and temporal offsets. Furthermore, we use the raw, non-normalized values of the underlying metrics (\textit{e.g.}, per-packet delays) to ensure our characterization reflects the natural scale and distribution of events in the network.

\subsection{Flow Overlap Quantification}
\begin{figure}[htbp]
    \centering
    \includegraphics[width=.8\columnwidth, trim=0 .5cm 0 .8cm,clip]{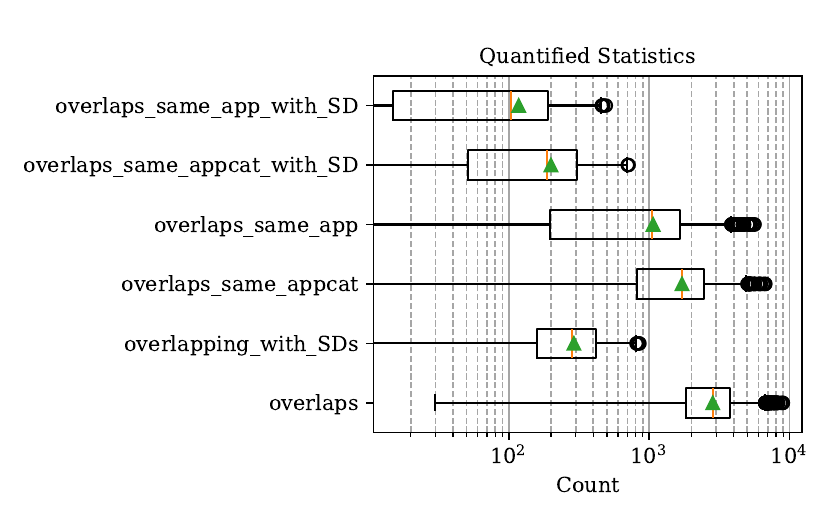}
    \caption{Distribution of covering flow counts per target flow. The plot shows a high volume of general overlaps but a much lower count for overlaps that also contain SD events, highlighting a signal sparsity challenge.}
    \label{fig:stats_quantified}
\end{figure}

First, we sought to understand the volume of potential data available for correlation. \Cref{fig:stats_quantified} shows the distribution of the number of covering flows per target flow. The opportunity for correlation is immense: a typical target flow is overlapped by thousands of concurrent flows, with a mean of 2836.2. This confirms that the raw material for inter-flow analysis exists in abundance.

However, the figure also reveals a critical challenge: \textit{sparsity of the most explicit signals}. When we filter these covering flows to count only those that also contain a SD event, the mean count drops by an order of magnitude to 290.9. This finding indicates that while the network is dense with activity, explicit degradation events are far less common, suggesting that a successful model cannot rely solely on finding co-occurring SDs.

\subsection{Timeliness of Correlated Signals}
\begin{figure}[htbp]
    \centering
    \includegraphics[width=.8\columnwidth, trim=0 .3cm 0 .6cm,clip]{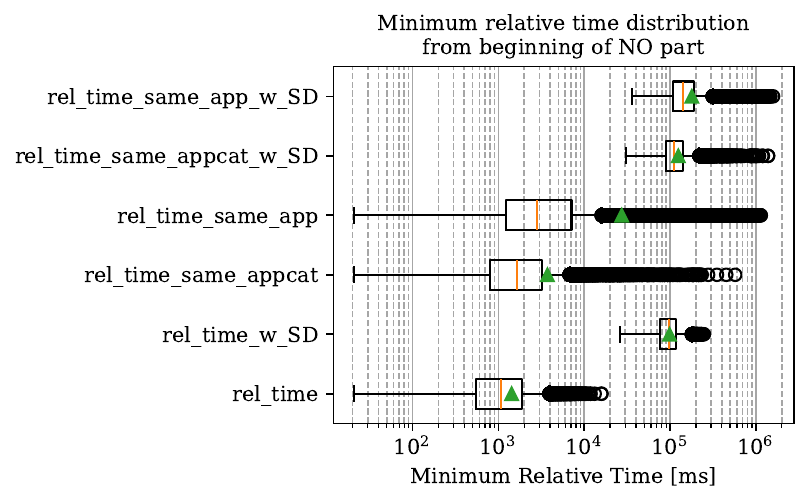}
    \caption{Distribution of time to the first covering flow. While generic overlaps appear quickly, the first SD-containing overlap is often significantly delayed.}
    \label{fig:stats_min_time}
\end{figure}

Next, we analyzed how quickly this inter-flow information becomes available. \Cref{fig:stats_min_time} displays the distribution of the time delay from a target flow's O/NO split to the start of the first covering flow. A generic covering flow appears almost immediately, with a mean delay of just 1.43 seconds. This confirms that some form of contextual data is available in near real-time.

The challenge, however, is again revealed when focusing on the most valuable signals. The time to the first covering flow that contains an SD event is significantly longer, with a mean of 98 seconds. For many target flows, which may only last a few minutes, this critical signal could arrive too late to be useful for real-time intervention. This finding strongly motivates the need for a modeling approach that can extract predictive information from all available covering flows, not just the rare ones that also exhibit explicit SD.

\subsection{Temporal Alignment of SD Events}

Despite the challenges of sparsity and delay, our most critical analysis validates the framework's potential for real-time detection by examining the optimal temporal alignment of co-occurring SD events. To quantify this, for each target flow containing one or more SD events, we identified all SD events across all of its covering flows. We then computed the temporal distance between the center-point of every possible target-SD/covering-SD pair. From this set of all pair-wise distances for a given target flow, we selected the single \textit{minimum value} to represent the ``best temporal alignment'' for that flow. This method ensures we capture the most optimistic case of correlation between any two degradation events.

\begin{figure}[htbp]
    \centering
    \includegraphics[width=.8\columnwidth, trim=0 .2cm 0 .2cm,clip]{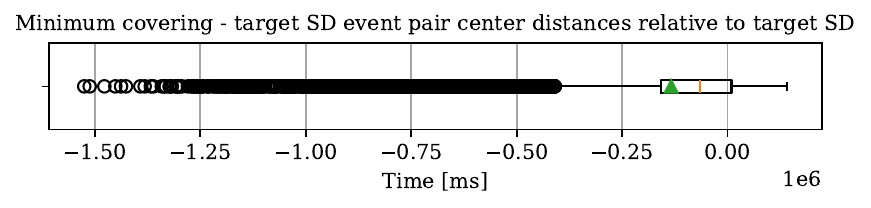}
    \caption{Distribution of the temporal distance between the center of a target SD event and the center of its best-aligned covering SD event. The negative skew indicates that the covering SD event often precedes the target SD.}
    \label{fig:stats_center_dist}
\end{figure}

As shown in \Cref{fig:stats_center_dist}, the distribution of these minimum-distance values is notably skewed negative, with a median offset of -66 seconds. This is a powerful finding: it demonstrates that the SD event in a covering flow often \textit{precedes} the event in the target flow, even under the most optimistic alignment. This confirms that the signal in concurrent flows is not just correlated but is often predictive, offering a window of opportunity for timely, or even pre-emptive, detection. This result provides the core empirical justification for our inter-flow framework and motivates the development of a detection model to harness this predictive potential.

\section{Experimental Evaluation of an Inter-Flow Model}
\label{sec:evaluation}

The statistical analysis in \Cref{sec:inter_flow_stats} confirmed that while the data for inter-flow correlation is abundant and holds predictive potential, it is also complex, sparse, and temporally misaligned. Guided by these insights, we designed an experiment to test a specific, pragmatic implementation of our framework. The goal was to assess if a standard machine learning model could harness this challenging data, establishing a performance benchmark and uncovering the practical limitations of a straightforward approach.

\subsection{Experimental Setup}
For this initial feasibility study, we made a key design choice to create a tractable and consistent feature space for our models. This choice was directly motivated by the findings of our statistical analysis.

The experiments leverage the large-scale dataset from our prior work~\cite{bicski2024unveiling}, which was captured from multiple residential gateways over 5 consecutive days. The ground truth labels, indicating the presence of SD events, were generated using the robust Z-score and IQR-based analysis established in that work. For this study, we created a balanced dataset of 100,000 flows for training and a separate 100,000 for testing.

\begin{itemize}
    \item \textit{Flow Selection:} While our general framework (\Cref{sec:methodology}) can accommodate all covering flows, for this experiment we constrained the input to manage the high dimensionality revealed in our analysis. For each target flow, we selected only the \textit{first 30 covering flows that were fully contained} within the correlation window. This provides a consistent input vector for the model but knowingly accepts a trade-off, discarding potentially useful data from partial overlaps and later-arriving flows.
    \item \textit{Feature Vector Construction:} 
    % The input for our models was a single, wide feature vector created by \textit{concatenating} the feature set of the target flow with the corresponding feature sets from each of its 30 selected covering flows. This included intra-flow metrics (\textit{e.g.}, delay and jitter statistics) and, for each covering flow, its analogous observable features and its start time relative to the beginning of the correlation window.
    The model input was a single, wide feature vector formed by \textit{concatenating} the target flow’s feature set with those of its 30 selected covering flows. The intra-flow set comprised 17 features—delay, jitter, observable SD event statistics, and application/connection type (categorical features one-hot encoded)—plus a dynamic set of delay and jitter metrics based on the O/NO split. Each covering flow contributed the same raw metrics and its relative start time, excluding one-hot encoded application/connection type features. The resulting vector contained $14+2m+4_{\textit{OH encoded}}+30\cdot(14+2m)$ features, plus covering counts per application type to capture distribution without per-flow one-hot encoding. This yielded 948 features for O/NO split 5 and 1248 for split 10. Detailed cardinalities are provided in the digital artifact~\cite{github}.
    
    % \ap{25?? metrics capturing raw delay statistics, jitter, and packet size distributions. The feature set for each of the 30 covering flows consisted of these same raw, non-normalized metrics plus its relative start time, \textit{creating a final concatenated feature vector with 775?? features (25?? for the target + 30 * (XY+Z)??? for the covering flows).}}
\end{itemize}

We trained three primary models: a regularized Logistic Regression, a feed-forward Multi-Layer Perceptron (MLP), and a gradient-boosted tree model using the XGBoost library. All models were optimized using a 5-fold cross-validated Grid Search to find the best hyperparameters. Standard scaling was applied on the input features of MLP. This experimental setup was designed to answer a crucial question: can a model architecture that performed well on the simpler, single-flow problem of our previous work~\cite{bicski2024early} adapt to learn from this much more complex, high-dimensional inter-flow feature set?

\subsection{Results and Discussion}
We trained several models for two primary tasks: classifying the presence of a SD event, and regressing its specific characteristics. The results present a nuanced picture of success and failure.

\begin{figure*}[htbp]
    \centering
    \subcaptionbox{ROC Curves show high AUROC for XGBoost.\label{fig:roc_curves}}{%
        \includegraphics[width=0.48\textwidth]{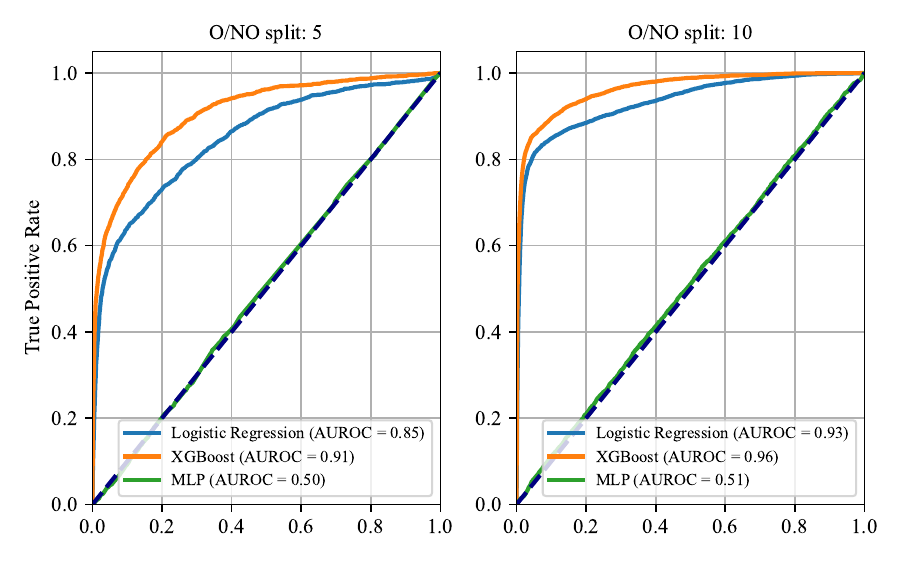}
    }
    \hfill
    \subcaptionbox{Classification metrics for O/NO=10 split.\label{fig:class_metrics}}{%
        \includegraphics[width=0.48\textwidth]{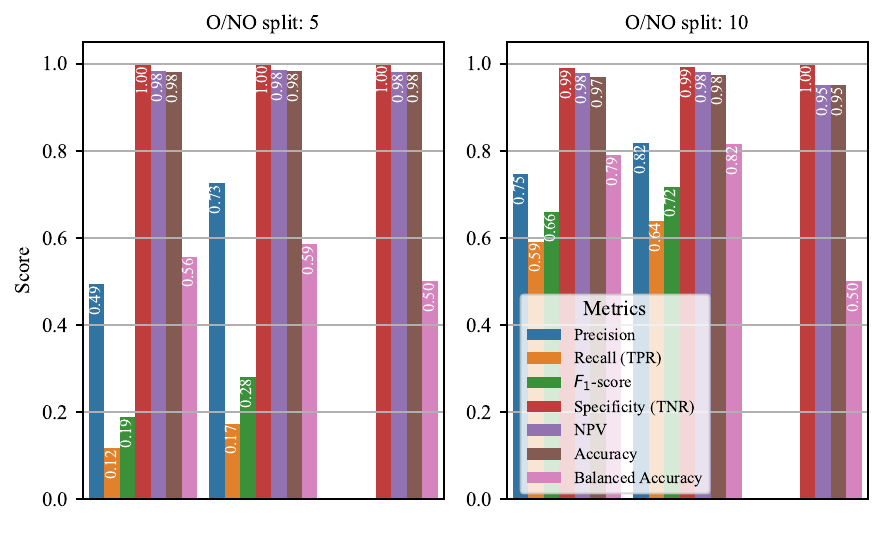}
    }
    \caption{Classification performance for the inter-flow detection model. While the top-line metrics such as AUROC and Balanced Accuracy appear strong, the overall performance did not show an improvement over purely intra-flow models, providing the first hint of the model's limitations.}
    \label{fig:inter-flow-classification}
\end{figure*}

For the classification task, the XGBoost model proved to be the most effective. As shown in \Cref{fig:inter-flow-classification}, it achieved a high AUROC of 0.96 and a strong balanced accuracy of 0.82 for the O/NO split of 10. On the surface, these strong top-line metrics suggest that the inter-flow model is viable and successful.

However, this initial success is misleading. The model’s performance collapsed on more granular regression tasks. As detailed in \Cref{tab:inter-flow_regression}, the errors for predicting SD event characteristics like start and end times were on the order of minutes, with high MAPE values. This stark contrast between classification success and regression failure strongly indicated that the model was not learning the underlying temporal dynamics of the inter-flow data.

\begin{table}[htbp]
    \centering
    \caption{Regression metrics for the XGBoost model on flows containing SD events (O/NO=10).}
    \label{tab:inter-flow_regression}
    \resizebox{\columnwidth}{!}{%
    \begin{tabular}{llccc}
    \toprule
    \textbf{Target Metric} & \textbf{MAE (s)} & \textbf{RMSE (s)} & \textbf{MAPE} \\
    \midrule
    \multirow{1}{*}{SD count} & 0.48 & 0.68 & 0.42 \\
    \multirow{1}{*}{Longest SD length (s)} & 207.0 & 309.6 & 0.79 \\
    \multirow{1}{*}{Longest SD start (s)} & 98.7 & 171.4 & 0.47 \\
    \multirow{1}{*}{Longest SD end (s)} & 282.3 & 406.2 & 0.46 \\
    \bottomrule
    \end{tabular}
    }
\end{table}

\begin{figure}[htbp]
    \centering
    \includegraphics[width=\columnwidth, trim=0 .3cm 0 .6cm,clip]{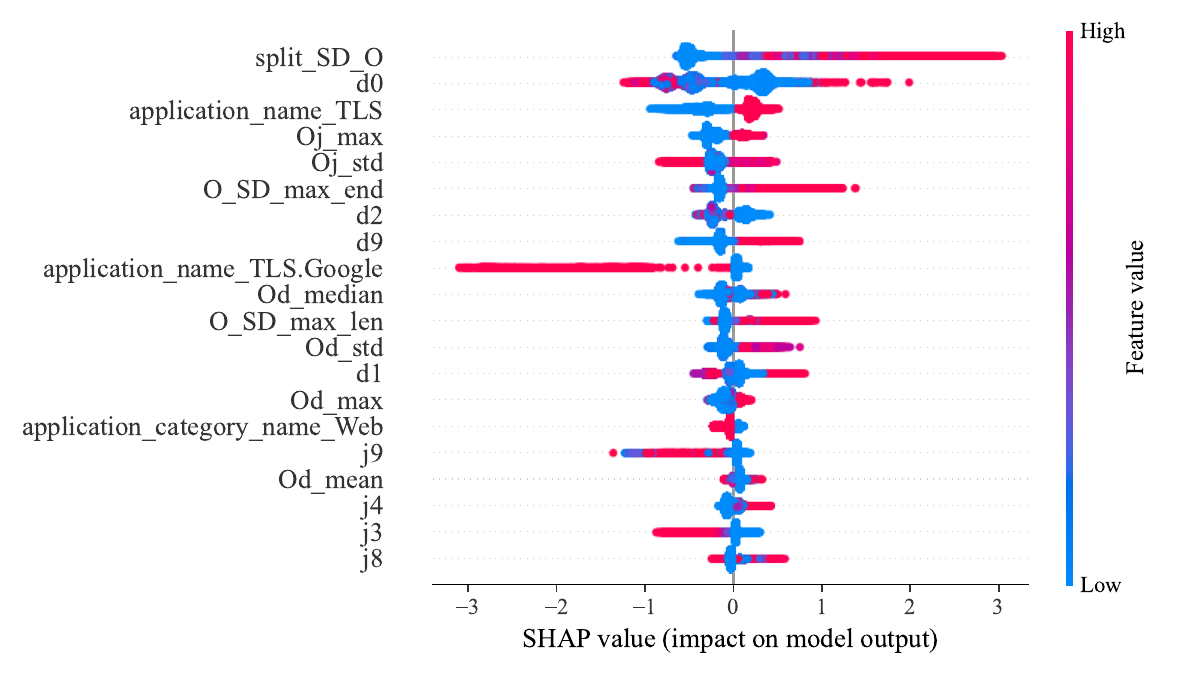}
    \caption{SHAP analysis of the classification model trained for O/NO split 10. Features with \texttt{d} prefix relate to delay measurements, while those with \texttt{j} are jitter related. \texttt{split\_SD\_O} shows how close are the delay and jitter measurements at the end of the O part to becoming an SD event. Features from covering flows would have the \texttt{cov\_} prefix on them.}
    \label{fig:shap}
\end{figure}

The definitive insight came from a \textit{feature importance analysis} of the successful XGBoost classification model,
conducted using SHAP (SHapley Additive exPlanations) shown in \Cref{fig:shap}.
Despite being provided with hundreds of inter-flow features from the 30 covering flows, the model learned to almost completely \textit{ignore them}. Its predictions were overwhelmingly based on the intra-flow features of the target flow itself.

This finding reveals a fundamental mismatch between the problem's complexity and the model's capabilities. By concatenating features, we created an extremely high-dimensional and sparse feature space. A model like XGBoost, which builds decisions by splitting on individual features, struggles to find meaningful relationships across hundreds of weakly correlated inputs. Furthermore, this simple vector provides no structural or temporal information; from the model's perspective, the delay metrics of flow \#3 are completely independent of the metrics of flow \#4. Lacking this relational context, the model defaults to the features it understands: the strong, familiar, and lower-dimensional signals from the target flow itself.

This is the central finding of our experimental evaluation: 
the model achieved good classification scores by simply replicating the performance of the intra-flow model using the target flow's own data, and it failed at regression because it never learned from the rich contextual data that could have helped characterize the events. The same model architecture that succeeded on the simpler intra-flow problem failed to adapt to the more complex inter-flow task. This does not invalidate our framework's principle, but it proves that a naive feature representation---in this case, simple concatenation---is insufficient for a standard model to learn from the sparse, high-dimensional, and time-shifted context of concurrent flows. This experiment successfully reveals that the primary challenge of inter-flow analysis lies not in the data's existence, but in its effective representation and modeling.

This finding provides a clear roadmap for future research. The logical path forward is to explore architectures inherently designed for relational and temporal data. Graph Neural Networks (GNNs) are particularly well-suited to this problem. A GNN could represent each flow as a node in a dynamic graph, with node attributes defined by the flow's observable features and edges representing temporal proximity or application similarity. Through message passing, a GNN could learn a rich, contextual representation of the network's local state, directly addressing the limitations of the simpler model. Furthermore, while this study was limited to temporal correlations, future work should explore spatial correlation by building graphs that connect flows across multiple network vantage points, a step toward capturing the localized nature of some degradation events.

\section{Conclusion and Future Work}
\label{sec:conclusion}

In this paper, we introduced a novel inter-flow correlation framework to address monitoring blind spots caused by hardware acceleration. Our primary contribution is a dual finding that bridges theory and practice. First, a comprehensive statistical analysis validated our framework's core premise: a predictive signal for SD does exist in concurrent network flows. Second, our experimental evaluation demonstrated that harnessing this signal is non-trivial. A standard machine learning model, despite its success in simpler intra-flow contexts, failed to leverage the complex inter-flow data when it was presented via simple feature concatenation.

% This result crystallizes the true challenge of inter-flow analysis: it is not a data availability problem, but one of feature representation and modeling complexity. Therefore, our work provides a clear outline for future research. The logical and most promising path forward is to move beyond simple concatenation and explore architectures designed for relational and temporal data. Graph Neural Networks, capable of learning from the graph structure of concurrent flows, and attention-based models represent the necessary next step to unlock the full potential of this new monitoring paradigm.

This result crystallizes the true challenge of inter-flow analysis: it is not a data availability problem, but one of feature representation and modeling complexity. Our work therefore provides a foundational analysis and a clear, empirically-grounded justification for future research into the more sophisticated, structure-aware models required to solve this important problem.

\section*{Data and Code Availability}

% To ensure the reproducibility of our results and to provide access to supplementary materials, including the full set of statistical analyses and figures that go beyond what was presented in this paper, our complete source code and data are publicly available~\cite{github}.

To ensure the reproducibility of our results, all data, analysis scripts, and code for generating figures are publicly available at~\cite{github}. This includes supplementary analyses not presented in the main text.

\section*{Acknowledgment}

Supported by the János Bolyai Research Scholarship of the Hungarian Academy of Sciences.
This work has been part of Celtic-Next project RAI-6Green: Robust and AI Native 6G for Green Networks with project-id: C2023/1-9 funded by 2024-1.2.6-EUREKA-2024-00009.
This work was also supported by projects TKP2021-NVA-02 (financed under the TKP2021-NVA scheme) and 2024-1.2.6-EUREKA-2024-00009 (financed under the 2024-1.2.6-EUREKA scheme), implemented with support provided by the Ministry of Culture and Innovation of Hungary from the National Research, Development and Innovation Fund.

\printbibliography

\end{document}